\documentstyle[PASJadd,graphicx]{PASJ95}
\makeatletter
\def\thefigure{\@arabic\c@figure}
\def\thetable{\@arabic\c@table}
\makeatother
\begin{document}

%
% Title and Authors, ...
%

\title{%
Evolution of Clusters of Galaxies II: Dependence on Initial Cluster Model}

\author{%
Tomohiro {\scshape Sensui},$^1$ Yoko {\scshape Funato},$^2$ and Junichiro {\scshape Makino}$^1$\\[1ex]
$^1$ {\itshape Department of Astronomy, The University of Tokyo, Bunkyo-ku, Tokyo 113-0033}\\
{\itshape E-mail(TS): sensui@astron.s.u-tokyo.ac.jp}\\
$^2$ {\itshape General Systems Sciences, Graduate Division of International and Interdisciplinary Studies,}\\
{\itshape The University of Tokyo, Meguro-ku, Tokyo 153-8902}}

%
% Abstract and Keywords
%

\abst{%
We investigated the evolution of clusters of galaxies using self-consistent $N$-body simulations.
We varied the initial model of galaxies and clusters, and studied the dependence of evolution on initial conditions.
We found that the growth rate of the common halo depends only weakly on galaxy models.
On the other hand, the growth rate depends strongly on cluster models.
Initially the growth rate is higher for cluster models with higher central density.
However, this high growth rate drops in a few crossing times of the cluster, and after several crossing times, roughly half of the total mass is in the common halo in all models we considered.
In the central region of clusters density cusps with the profile $\rho \sim r^{-1.2}$ develops regardless of the models of clusters and galaxies.
We also found that the galaxies evolved so as to satisfy the relation between the masses of galaxies $m_{\mathrm{gx}}$ and their velocity dispersion $\sigma_{\mathrm{gx}}$ expressed as $m_{\mathrm{gx}}\propto\sigma_{\mathrm{gx}}^{3\sim 4}$ for all galaxy models as a consequence of their dynamical evolution through galaxy--galaxy interactions.
We discuss the relation between our result and the observed Faber--Jackson relation.}

\kword{%
galaxies: clusters: general ---
galaxies: evolution ---
galaxies: interactions}

\maketitle
\thispagestyle{headings}

%
% Introduction
%

% section 1
\section{Introduction}

In recent years, observations of structures of galaxies in clusters at high-$z$ ($z\sim 1$) have become possible due to the advances in telescopes, related instruments and methods of analysis.
Such observations of high-$z$ clusters allow us to study the evolution of clusters and galaxies in clusters by comparing high- and low-$z$ clusters.
Using an {\sl I}-band magnitude-selected sample of 81 confirmed cluster members covered by HST WFPC2, van Dokkum et al.\ (1999) reported that the fraction of mergers in clusters is high in high-$z$ region and the fraction drops quickly in low-$z$ clusters.
In the field, merger fraction seems to drop more slowly than that in clusters.
Their result suggests that mergings of galaxies take place mainly in the formation epoch of a cluster.
After the cluster virialized, or, in the region where the galaxies have virialized, mergings of galaxies become rare.

However, merging is not the only way for galaxies in clusters to evolve.
Ellipticals and S0 galaxies are much more abundant in clusters than those in the field.
They are more abundant in clusters in low-$z$ than in high-$z$ (e.g., Couch et al.\ 1998) and, especially, in central high density regions in such clusters (density-morphology relation, Dressler 1980).
These observations suggest that galaxies dynamically evolve after their parent cluster virialized.

In this paper, we investigate how clusters of galaxies and their member galaxies evolve after the formation of clusters themselves.

$N$-body simulation is a powerful tool to study the formation and evolution of clusters of galaxies.
Two types of $N$-body simulations of clusters of galaxies have been used.

One is cosmological $N$-body simulations, in which the formation process of galaxies and clusters is simulated in the general framework of CDM cosmologies.
Many cosmological simulations have been carried out to study how clusters of galaxies were formed from initial density fluctuations (Bertschinger 1998 for a review) and to determine the ``best fit'' values of cosmological parameters such as $\Omega$, $H_{0}$, $\Lambda$, and so on.

Due to the advances in computer technologies and algorithms, the number of particles in simulations have significantly increased from $\sim 10^{3}$ in the late 1970s (Miyoshi and Kihara 1975; Fall 1978; Aarseth and Fall 1980) to $\sim 10^{6}$ (Moore et al.\ 1998; Ghigna et al.\ 1998; Okamoto and Habe 1999) at present.
In 1980s, galaxies are treated as point particles in cosmological simulations of clusters.
However, recently there are some studies in which both the formation of a cluster and that of galaxies in it are followed self-consistently in a single large simulation.
In these simulations, it is expected that a cluster has a hierarchical structure and small clumps of particles can be regarded as galactic halos formed in the clusters.
They tried to compare properties of ``galactic halos'' such as density profiles or velocity dispersions with those of observed cluster galaxies.

However, the cosmological simulation is not a universal tool which allows us to study all aspects of evolution of clusters.
In cosmological $N$-body simulations, dynamical processes with different spatial and time scales take place simultaneously.
For example, galaxies in the central virialized region of a cluster collide with each other in high speed.
As a result, they decrease their masses due to stripping.
At the same time, newly formed galaxies fall into the cluster, and some of infalling galaxies might merge with each other because of low relative velocity.
It is, therefore, difficult to understand the essential physical mechanisms of the formation and evolution of the structure of clusters in these simulations.
Even if these simulations reproduce the present-day structure, it is not clear how the present-day structure is formed.
Without the understanding of essential physical mechanisms, it is difficult to understand what are the common properties of clusters and galaxies and why they are.

The reliability of simulations is also a problem.
Even if the number of particles is large enough to express the overall structure of a typical galactic halo on average, it is still far from sufficient to resolve dwarfs, galaxies in central regions of clusters, or central regions of typical galaxies.

In order to understand how galaxies evolve after virialization of the parent cluster, it is helpful to perform the other type of simulations in which we simulate the evolution of model clusters of galaxies (e.g., Funato et al.\ 1993; Bode et al.\ 1994; Garijo et al.\ 1997; Sensui et al.\ 1999 (paper I)).

By constructing the model clusters and model galaxies appropriately as initial conditions, we can quantitatively and separately study the effect of each dynamical process, such as mass stripping, merging, dynamical friction, tidal disruption due to the mean field of the parent clusters and so on.
Using understanding of these individual dynamical processes obtained from model cluster simulations, we can improve interpretation of the result of cosmological simulations.
Understanding these individual dynamical processes obtained from model cluster simulations and results of cosmological simulations complementarily help us to understand evolution of galaxies and clusters.

This paper is the second of the series of papers in which we study the evolution of model clusters.
In our study (paper I and present paper) we set the initial condition of cluster models so that the effect of galaxy--galaxy interaction is clearly visible since interactions seem to be the dominant mechanism for the dynamical evolution of clusters of galaxies.
We obtained quantitative results on how the mass of galaxies, and the structure of individual galaxies and that of a cluster, evolve.

In paper I, we used a Plummer model for both a galaxy model and a cluster model.
We found that more than half of the total mass escaped from individual galaxies and that a common halo was formed.
The growth rate of the common halo depended on the size of the cluster only weakly.
We also found that the density profile of the cluster is expressed as $\rho \sim r^{-1.2}$ in the central region.

We, however, cannot draw a general conclusion from simulations with the Plummer model only.
Since frequency of the galaxy--galaxy interaction is affected by distribution of galaxies within a cluster, the global structure of the cluster, such as a core radius or a density profile, may affect the results.
Furthermore, the Plummer model is not appropriate as a model of a real galaxy or a real cluster, as this model has a core much larger than those of observed galaxies or clusters.
Therefore, studies with other models for galaxies and clusters are necessary to understand the evolution of real clusters.

In this study, we adopted a Plummer model, King models and a Hernquist model (Hernquist 1990) as models of galaxies and clusters to study how the initial profiles of the cluster and galaxies affect the result.
By performing the simulations for different model galaxies and clusters with different density structures, we can study the effect of the initial model on the result systematically.
We investigated the growth timescale of the common halo, the density profile of clusters, and the evolution of the mass and structure of galaxies.

In section~2 we describe initial conditions which we used in our simulations.
Time integration method and galaxy identification method are described in section~3.
In section~4 we present the results.
A summary and discussions are given in section~5.

%
% Initial Conditions
%

% section 2
\section{Initial Conditions}
% subsection 2.1
\subsection{Units}

We used a system of units in which $m=G=1$ and $e=-1/4$, where $G$ is the gravitational constant and $m$ and $e$ are the mass and the energy of one initial galaxy (Heggie and Mathieu 1986).
Assuming that the mass unit corresponds to $10^{12}M_{\solar}$ and the length unit to 30 kpc, we find that the time unit corresponds to 110 Myr.

% subsection 2.2
\subsection{Initial Models of Galaxies and Clusters}

\begin{table}[t]\small
 \caption{Initial conditions.}
 \begin{center}
  \begin{tabular}{lcc}
   \hline\hline
   \multicolumn{1}{c}{ID} & galaxy model & cluster model\\
   \hline
   PP  & Plummer & Plummer \\
   PK3 & Plummer & King~3 \\
   PK5 & Plummer & King~5 \\
   PK7 & Plummer & King~7 \\
   PH  & Plummer & Hernquist \\
   K3P  & King~3 & Plummer \\
   K7P  & King~7 & Plummer \\
   K9P  & King~9 & Plummer \\
   HP  & Hernquist & Plummer \\
   \hline
  \end{tabular}
 \end{center}
\end{table}

For models of galaxies, we adopted a Plummer model, King models whose non-dimensional potential depth are 3, 7, and 9, and a Hernquist model.
For all galaxy models, the mass $m$ and the virial radius $r_{0}$ are 1, and the velocity dispersion $\sigma_{\mathrm{gx}}$ is $1/\sqrt{2}$.
All galaxy models comprise 2048 particles, so the mass of a particle is $1/2048$.
As we showed in paper I, this number of particles per one galaxy is sufficiently large to suppress the two-body relaxation, as far as the evolution of the masses of the individual galaxies and the structure of the common halo are concerned.

We adopted a Plummer model, King models whose non-dimensional potential depth are 3, 5, and 7, and a Hernquist model as cluster models.
A King model with a deeper central potential (and a high central concentration) gives a better fitting for a real cluster than a Plummer model.

For all cluster models, the number of galaxies is 128, and all of the mass was initially attached to galaxies.
The total mass $M$ of a cluster is, therefore, 128 in all runs.
The virial radius of the cluster, $R_{\mathrm{vr}}$, is 20.
The velocity dispersion $\sigma_{\mathrm{cl}}$, the total energy $E$, and the crossing time $T_{\mathrm{cr}}$ are:
%
% equations 1,2,3
\begin{eqnarray}
 \sigma_{\mathrm{cl}}
  &=& 4/\sqrt{5} = 1.789,\\
 E
  &=& \frac{1}{2}M{\sigma_{\mathrm{cl}}}^{2}-\frac{M^{2}}{2R_{\mathrm{vr}}}
   = -204.8,\\
 T_{\mathrm{cr}}
  &=& \frac{2R_{\mathrm{vr}}}{\sigma_{\mathrm{cl}}}
   = GM^{5/2}|2E|^{-3/2} = 10\sqrt{5}.
\end{eqnarray}
Under the same conversion factors as used for galaxies, the total mass of the cluster corresponds to $1.28\times 10^{14}M_{\solar}$.
The velocity dispersion of galaxies in the cluster is 480 km/sec, the virial radius of the cluster is 0.6 Mpc and the crossing time is 2.5 Gyr.

In table~1, we summarize the initial models which we used.
The suffixes of King models denote the non-dimensional potential depth of King models, $\Psi(0)/\sigma^{2}$, where $\Psi(0)$ is the central potential of the model, and $\sigma^{2}$ is the mean one-dimensional velocity dispersion of the distribution function of the King model (King 1966).

%
% Numerical Methods
%

% section 3
\section{Numerical Methods}
% subsection 3.1
\subsection{Time Integration}

The total number of particles $N$ is 262144.
In order to carry out calculations with this large number of particles fast, we used GRAPE-4 (Makino et al.\ 1997) and GRAPE-5 (Kawai et al.\ 2000) with the Barnes--Hut tree algorithm (Barnes and Hut 1986; Makino 1991; Athanassoula et al.\ 1998).
The softening parameter $\epsilon$ was 0.025, the opening angle $\theta$ was 0.75, and the time step $\Delta t$ was 1/64.
One time step took about 14 seconds with GRAPE-4 and 5 seconds with GRAPE-5.

In all runs, the time integration was carried out using the leap-frog method.
The errors in the total energy $\Delta E/E_{0}$ were of the order of $10^{-3}$ in all runs.

% subsection 3.2
\subsection{Galaxy Recognition}

The method of recognition of individual galaxies is the same as that used in paper I except for the treatment of mergers.
We added an automatic procedure to automatically recognize mergers which we will explain in section~3.3.

To study how galaxies evolve in a cluster, we need to determine which particles have escaped from them.
We define a galaxy as a self-gravitating bound object.
In practice, we calculated the number of escapers by the following procedure.

Initially, each particle belongs to one unique galaxy.
We call this galaxy as a parent galaxy.
At each time step, we calculated the binding energy of each particle in its parent galaxy.
If the binding energy was positive, we regarded that particle as having escaped from its parent galaxy.
The binding energy was calculated using those particles that were still bound to the parent galaxy.
Thus, we needed to iterate this calculation several times to stabilize the membership.
Using this algorithm, we could trace the identity of a galaxy, even after 90\% of the mass had escaped.
The limitation of the present method is that we cannot deal with the exchange of particles between galaxies.
Since the fraction of mass exchanged is expected to be negligible, neglecting the exchange would not affect our result.

Note that this procedure is applied to each galaxy, after possible mergers are identified.
Otherwise, a galaxy which merged with others might be regarded as completely disrupted.

\begin{figure*}[tp]
\begin{center}
 \includegraphics*[height=.22\textheight]{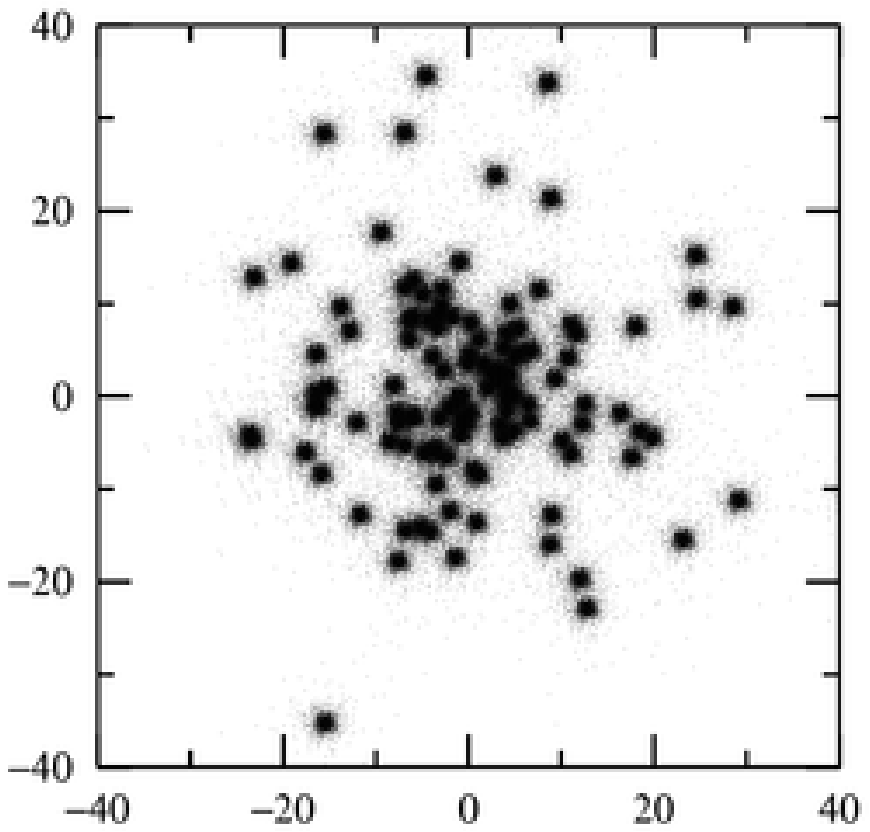}\hspace{2cm}
 \includegraphics*[height=.22\textheight]{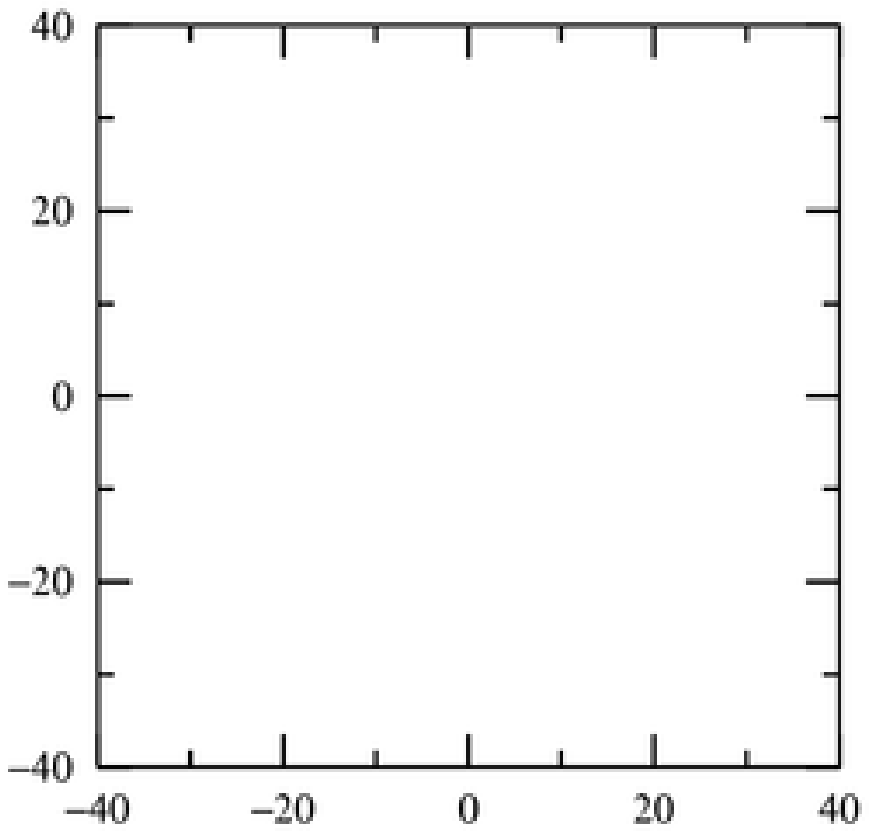}\\[-.4ex]
 \includegraphics*[height=.22\textheight]{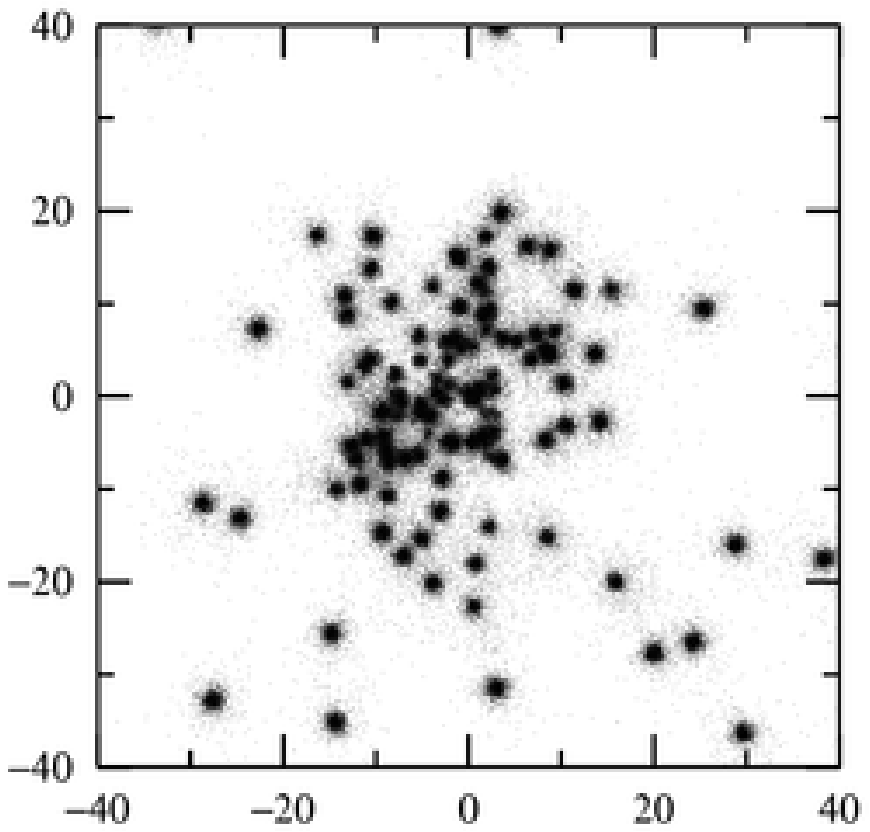}\hspace{2cm}
 \includegraphics*[height=.22\textheight]{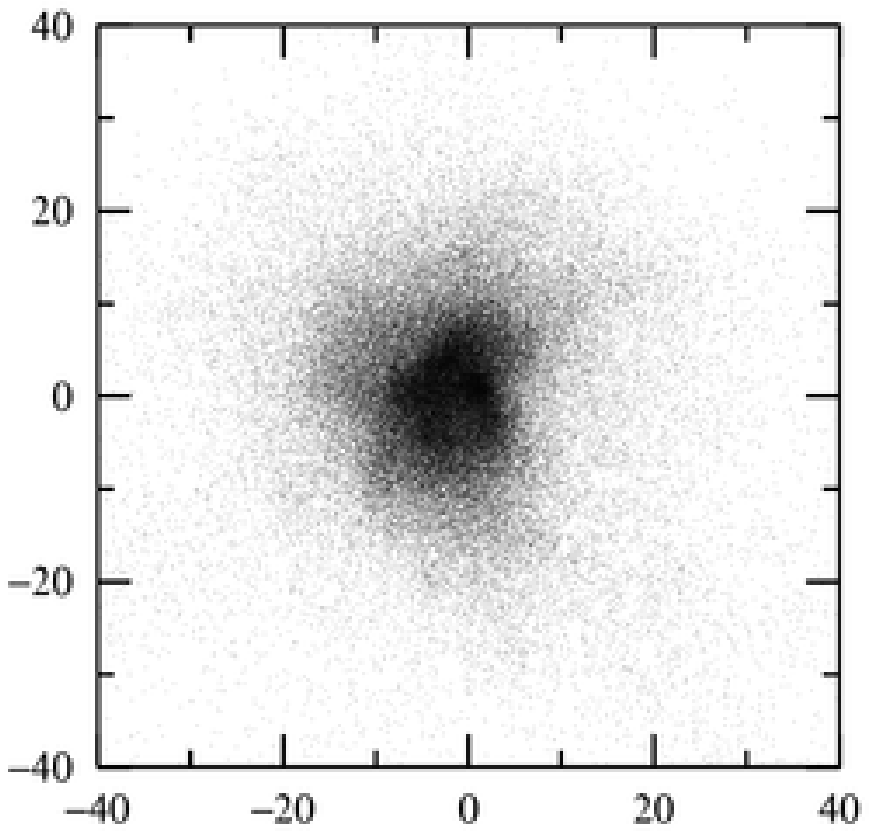}\\[-.4ex]
 \includegraphics*[height=.22\textheight]{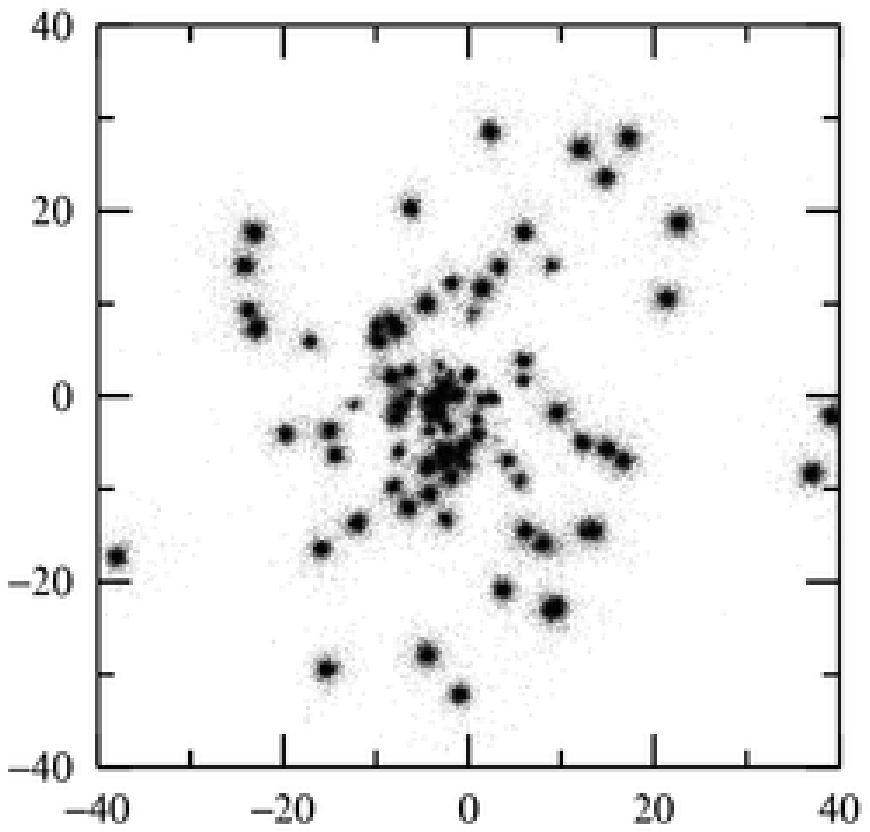}\hspace{2cm}
 \includegraphics*[height=.22\textheight]{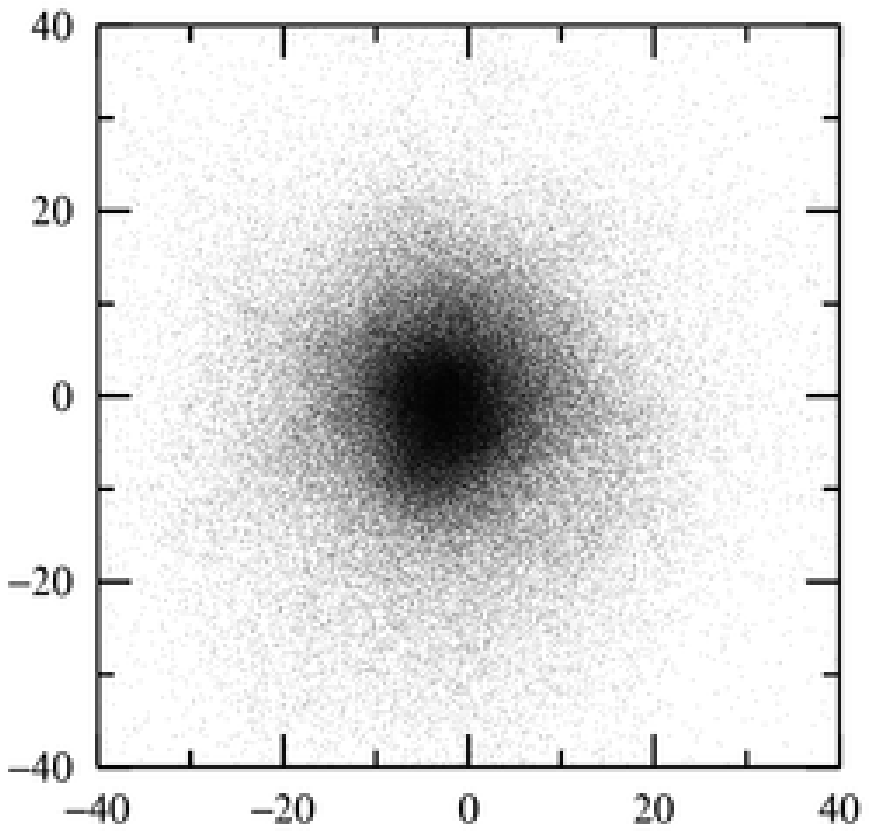}\\[-.4ex]
 \includegraphics*[height=.22\textheight]{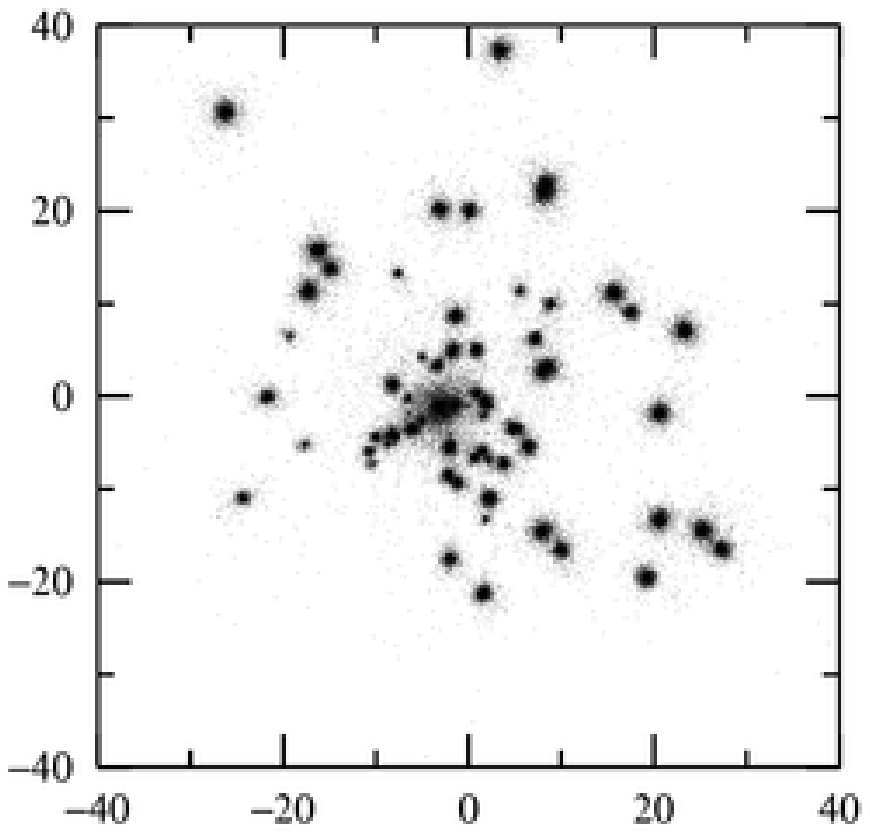}\hspace{2cm}
 \includegraphics*[height=.22\textheight]{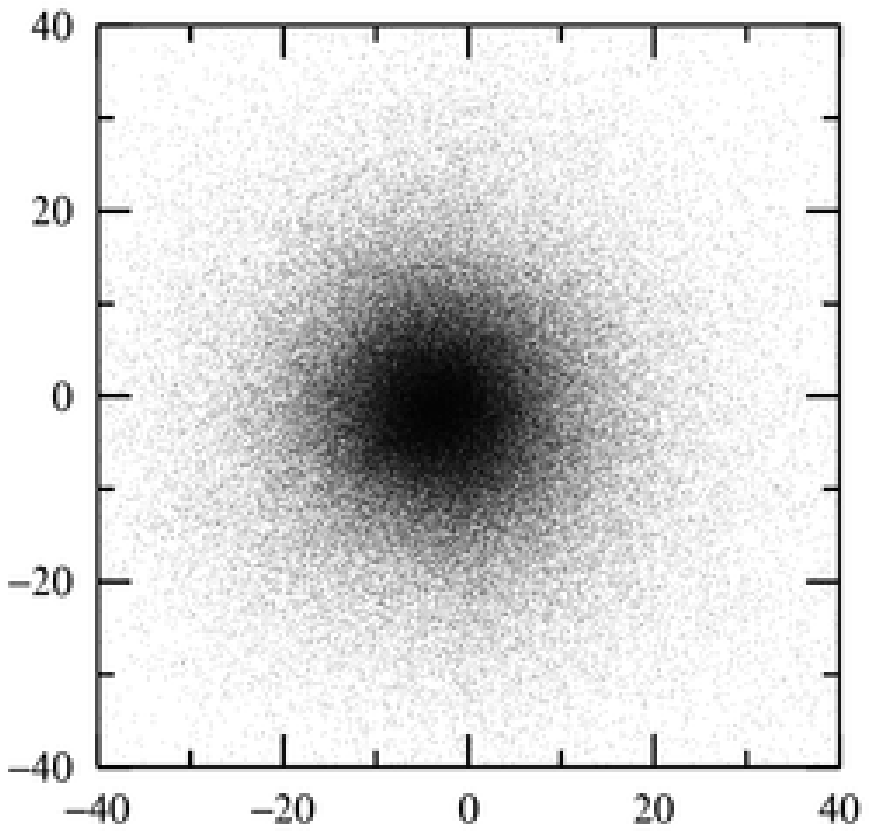}
\end{center}
\caption{%
Snapshots of the run PP.
Particles are projected onto the $x$--$y$ plane.
The left-hand side panels show the particles bound to individual galaxies.
The right-hand side panels show those escaped from galaxies to intracluster space, which formed a common halo.
Each row corresponds to $t=0$ (top), $2.2\, T_{\mathrm{cr}}$, $4.5\, T_{\mathrm{cr}}$, and $8.9\, T_{\mathrm{cr}}$ (bottom), respectively.}
\end{figure*}

% subsection 3.3
\subsection{Merger Identification}

To find merger candidates, we used the method similar to that used by Athanassoula et al.\ (1997).
We defined the center of a galaxy as
%
% equation 4
\begin{equation}
 \mbox{\boldmath$r$}_{\mathrm{c}}
  = \sum_{i=1}^{n}\rho_{i}\mbox{\boldmath$r$}_{i}\!
  \left/\, \sum_{i=1}^{n}\rho_{i} \right.,
\end{equation}
where $n$ is the number of particles in the galaxy, $\rho_{i}$ is the local density of particle $i$ calculated using the sixth nearest neighbor in the galaxy (Casertano and Hut 1985), and $\mbox{\boldmath$r$}_{i}$ is the position of particle $i$.
We calculated the radius of a galaxy as
%
% equation 5
\begin{equation}
 r_{\mathrm{gx}} =
  \sqrt{\sum_{i=1}^{n}\rho_{i}
  (\mbox{\boldmath$r$}_{i}-\mbox{\boldmath$r$}_{\mathrm{c}})^{2}\!
  \left/\, \displaystyle\sum_{i=1}^{n}\rho_{i} \right.}.
\end{equation}
We also calculated the central velocity dispersion of a galaxy $\varsigma_{\mathrm{gx}}$ by taking the root-mean-squared velocity of particles within $2\,r_{\mathrm{gx}}$ in the galaxy.

Using these two characteristic values, $r_{\mathrm{gx}}$ and $\varsigma_{\mathrm{gx}}$, we regard two galaxies as merged if the following criteria
%
% equations 6,7,8
\begin{eqnarray}
 R_{ij} &<& 0.5\min(r_{{\mathrm{gx}},i}\,, r_{{\mathrm{gx}},j})\\
 V_{ij} &<& 0.5\min(\varsigma_{{\mathrm{gx}},i}\,, \varsigma_{{\mathrm{gx}},j})\\
 r_{{\mathrm{gx}},i}\,, r_{{\mathrm{gx}},j} &<& r_{\mathrm{crit}}
\end{eqnarray}
are satisfied.
Here, $R_{ij}$ is the relative distance of two galaxies, $V_{ij}$ is the relative velocity of them, and $r_{\mathrm{crit}}$ is a criterion radius.
The first two criteria examine whether two galaxies are close enough to merge.

The last criterion examines whether they can be still recognized as galaxies or they have been already disrupted.
Athanassoula et al.\ (1997) did not use this last criterion because they were interested in the evolution of small groups of galaxies, and used only five galaxies for all runs.
On the other hand, we used 128 galaxies for all runs, and some galaxies were disrupted through their evolutions.
The particles which originally belonged to these disrupted galaxies are distributed throughout the whole cluster, as part of the common halo.
Thus, if we do not use the last criterion, these disrupted galaxies with large radius are identified as merged.
In other words, the common halo itself is recognized as a huge merger.

Physically speaking, when more than 50\% of the total mass goes to the common halo, it becomes a self-gravitational system in its own right, and it is correct that the common halo is recognized as ``a merger''.
However, here we regard mergers in a more conventional sense that the remnant of the merging events of several galaxies, and analyze the common halo as a distinct object, so we introduced the last criterion to exclude the common halo.
We set $r_{\mathrm{crit}}$ to 3 for runs K9P, PK7 and PH, and 5 for the others.

We carried out this check for all galaxy pairs, and if more than two galaxies merged to one galaxy, we regarded them as a multiple merger.

%
% Results
%

% section 4
\section{Results}
% section 4.1
\subsection{Snapshots}

Figure~1 shows snapshots for the run PP.
The panels in the left-hand side show particles bound to galaxies, and the panels in the right-hand side show particles escaped from galaxies to the intracluster space.
We can see from this figure that a common halo is formed and develops as particles escape, and that the growth timescale is of the order of the crossing time of the cluster (a few Gyrs).

In the following, we first investigate the growth timescale of the common halo and the structure of the whole cluster, and then the evolution of individual galaxies.

% subsection 4.2
\subsection{Properties of Clusters}
% subsubsection 4.2.1
\subsubsection{Evolution of common halos}

\begin{figure}[t]
\includegraphics*[width=.48\textwidth]{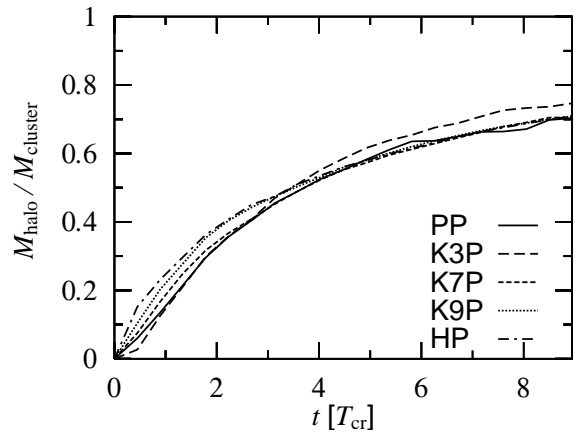}
\caption{%
Fractional mass of the common halo $M_{\mathrm{halo}}/M_{\mathrm{cluster}}$ plotted against time in unit of the cluster crossing time $T_{\mathrm{cr}}$, for runs PP (solid), K3P (long-dashed), K7P (short-dashed), K9P (dotted), and HP (dashed-dotted).}
\end{figure}

\begin{figure}[t]
\includegraphics*[width=.48\textwidth]{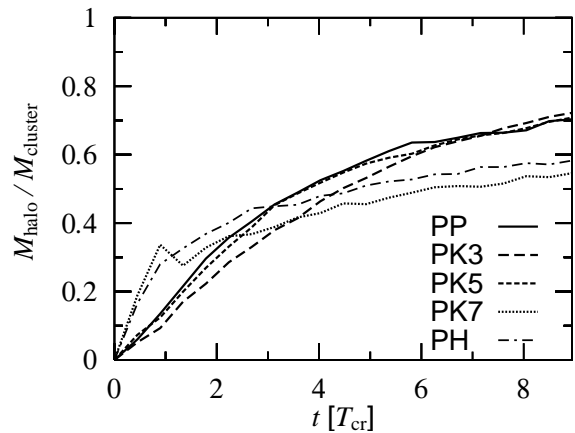}
\caption{%
Same as figure~2, but for runs PP (solid), PK3 (long-dashed), PK5 (short-dashed), PK7 (dotted), and PH (dashed-dotted).}
\end{figure}

Figure~2 shows the growth of mass in the common halo for runs with different galaxy mode
ls.
When the time is small, the growth rate of the common halo is slightly larger for runs with galaxy models with high central densities (e.g., K9P) than that for runs with galaxy models with low central densities (e.g., K3P).
It is understood as follows.
A galaxy with a higher central density also has more extended envelope than a galaxy with a low central density.
The mass in the outer part of a galaxy can escape easily from individual
 galaxies to the intracluster space through galaxy--galaxy interactions.
A galaxy with a more extended halo loses its mass more easily than that with a less extended halo.

After $t\sim 3\, T_{\mathrm{cr}}$, the growth rate of a common halo becomes almost the same for all runs except for the run K3P (but the difference is small).
In other words, after a large fraction of the mass in the outer part of a galaxy is lost, the growth rate of a common halo becomes almost independent of the internal structure of the initial galaxy model.

Figure~3 shows the evolution of the common halo for runs with different cluster models.
We can see two important results in figure~3.
The first one is that the growth rate of runs PK7 and PH are much faster than the others in early phase ($t\lsim 1\, T_{\mathrm{cr}}$), and the second one is that $M_{\mathrm{halo}}/M_{\mathrm{cluster}}\sim 0.5$ in all runs at $t\sim 4\, T_{\mathrm{cr}}$.
We will return to these issues in section~5.1.

For the run PK7, the fractional mass of the common halo decreases occasionally.
This is caused by our criterion to recognize merging events.
We sometimes found a massive virialized system which is actually the central region of the common halo, in particular for the runs with centrally concentrated cluster models.
As we described in section~3.3, this is physically correct since we are really looking at a compact self-gravitating object.
However, this does cause some large fluctuation on the mass of the common halo as we can see in figure~3.

\begin{figure}[t]
\includegraphics*[width=.48\textwidth]{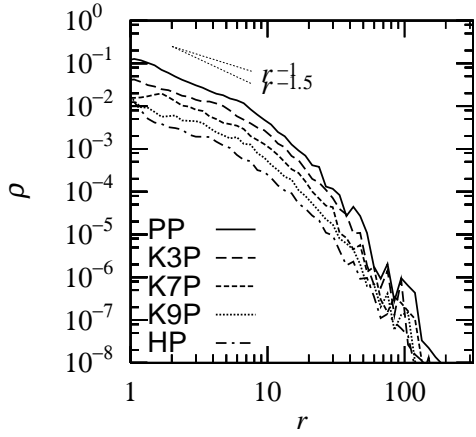}
\caption{%
The density profiles of clusters at $t=8.9\, T_{\mathrm{cr}}$ for runs PP (solid), K3P (long-dashed), K7P (short-dashed), K9P (dotted), and HP (dashed-dotted).
Densities are scaled to make them distinguishable from each other.}
\end{figure}

\begin{figure}[t]
(a)
\includegraphics*[width=.48\textwidth]{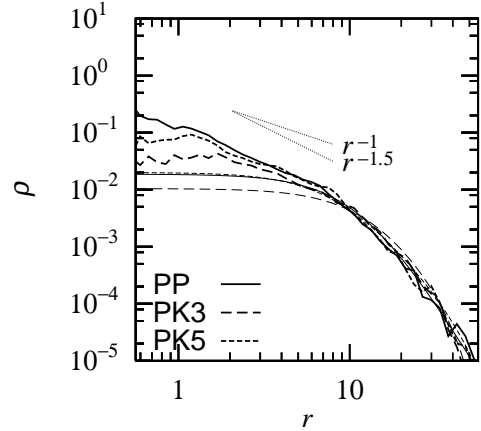}\\[-.4ex]
(b)
\includegraphics*[width=.48\textwidth]{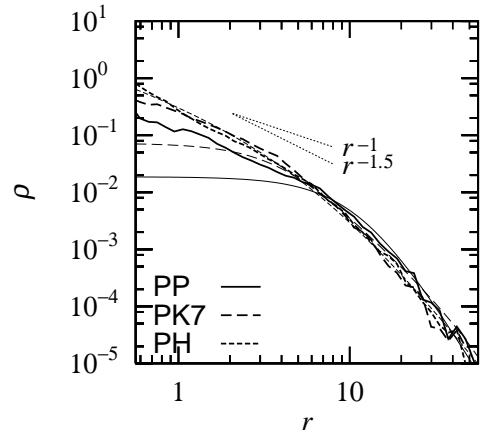}
\caption{%
Density profiles at central regions of clusters for (a) runs PP (solid), PK3 (long-dashed) and PK5 (short-dashed), and (b) runs PP (solid), PK7 (long-dashed) and PH (short-dashed).
Thin and thick curves correspond to the initial theoretical density profiles and profiles at $t=8.9\, T_{\mathrm{cr}}$, respectively.}
\end{figure}

% subsubsection 4.2.2
\subsubsection{Density profile}

Figure~4 shows the density profiles of clusters at $t=8.9\, T_{\mathrm{cr}}$ for runs in which the Plummer model is used as the initial cluster model.
Figure~4 shows that the difference in the initial galaxy models does not affect the density profiles of clusters.

Figure~5 shows the density profiles of core regions of clusters for runs with different cluster models.
Thin and thick curves correspond to the initial analytical density profiles and numerically obtained profiles at $t=8.9\, T_{\mathrm{cr}}$ for each model, respectively.

It is remarkable that in almost all runs the central density increases to realize $\rho \sim r^{-1.2}$ cusps, while the profile of outer region is practically unchanged.
To compare our results with other profiles proposed for clusters and galaxies with central cusps, we fitted our results using the Nuker law (Lauer et al.\ 1995; Byun et al.\ 1996),
%
% equation 9
\begin{equation}
 \hspace*{-5mm}
 \rho(r) = \rho_{\mathrm{b}} 2^{(\beta-\gamma)/\alpha}
  \left(\frac{r}{r_{\mathrm{b}}}\right)^{-\gamma}
  \left[1+\left(\frac{r}{r_{\mathrm{b}}}\right)^{\alpha}\right]^{-(\beta-\gamma)/\alpha}.
\end{equation}
Originally, Nuker law was used to fit surface brightness profiles of observed galaxies, but here we used it to fit density profiles.
A set of parameters $\alpha=1,\beta=4,\gamma=1$ corresponds to a Hernquist profile, and that of $\alpha=1,\beta=3,\gamma=1$ gives the universal profile (Navarro et al.\ 1997).

\begin{figure}[t]
(a)
\includegraphics*[width=.48\textwidth]{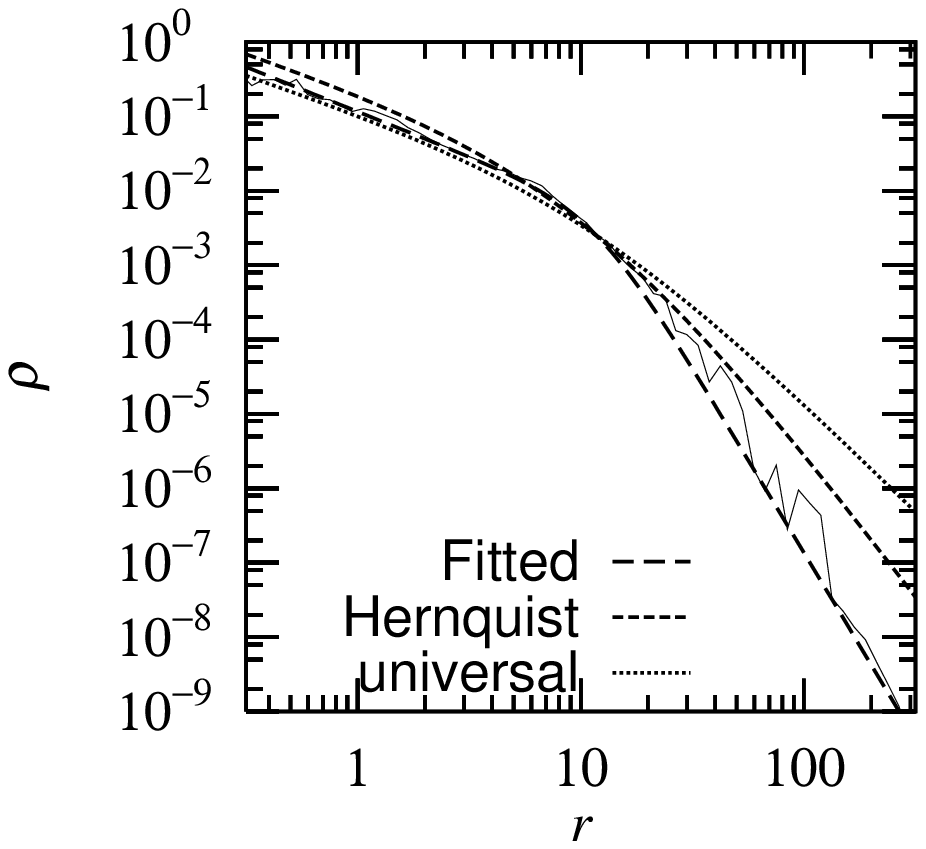}\\[-.4ex]
(b)
\includegraphics*[width=.48\textwidth]{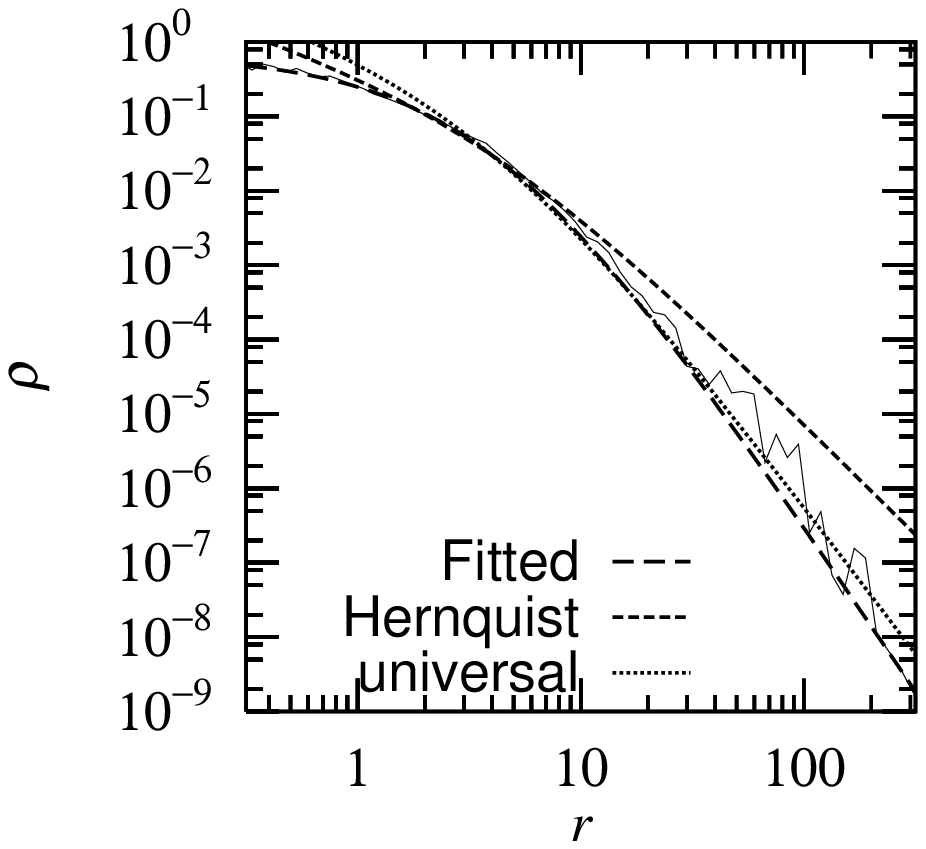}
\caption{%
Fractional mass of the common halo $M_{\mathrm{halo}}/M_{\mathrm{cluster}}$ plotted against time in unit of the cluster crossing time $T_{\mathrm{cr}}$, for runs PP (solid), K3P (long-dashed), K7P (short-dashed), K9P (dotted), and HP (dashed-dotted).}
\end{figure}

In figure~6, we plotted the ``best-fit'' profile as well as a Hernquist profile and the universal profile.
In each panel, we used the same $\rho_{\mathrm{b}}$ and $r_{\mathrm{b}}$ for all profiles.
Fitting parameters are shown in table~2.
The parameter set for the run PP is an example of typical profiles and that for the run PK7 is an exceptional one for the initial cluster model with very high central density.
Here, even though there is the central cusp region, the best-fit result says $\gamma=0$ simply because the cusp region is too narrow.

\begin{figure}[t]
\includegraphics*[width=.48\textwidth]{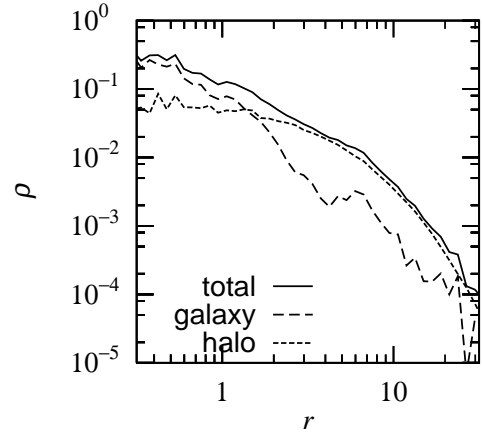}
\caption{%
Density profiles at the central region of the total cluster (solid), galaxies (long-dashed), and the halo (short-dashed) at $t=8.9\, T_{\mathrm{cr}}$ for the run PP.}
\end{figure}

\begin{table}[t]\small
 \caption{Fitting parameters in Nuker law.}
 \begin{center}
  \begin{tabular}{lccccc}
   \hline\hline
   \multicolumn{1}{c}{ID} & $\alpha$ & $\beta$ & $\gamma$ &
   $\rho_{\rm b}$ & $r_{\rm b}$ \\
   \hline
   PP  & 3 & 5 & 1.2 & 0.0025 & 11.78 \\
   PK7 & 1 & 4.5 & 0 & 0.03 & 4 \\[3ex]
   Hernquist & 1 & 3 & 1 & 0.0025 or 0.03 & 11.78 or 4 \\
   universal & 1 & 4 & 1 & 0.0025 or 0.03 & 11.78 or 4 \\
   \hline
  \end{tabular}
 \end{center}
\end{table}

Figure~7 shows the density profiles at the core region of the total cluster, galaxies, and the halo for the run PP at $t=8.9\, T_{\mathrm{cr}}$.
We can see that the galaxy component is dominant for the density profile at the core region of the cluster.
This tendency is observed in all models including PK7.

To investigate how the central density increases with time, we measured the ratio between the mass in galaxies and the total mass within given radii.
In figure~8 we show the result for three epochs for the run PP.
The relative mass in galaxies at the central region increases, even though the total mass in galaxies decreases.
This result implies that galaxies sink towards the center of the cluster due to the dynamical friction.
In other words, figure~8 suggests that the formation of the shallow cusps is caused by thermal relaxation of the cluster.

\begin{figure}[t]
\includegraphics*[width=.48\textwidth]{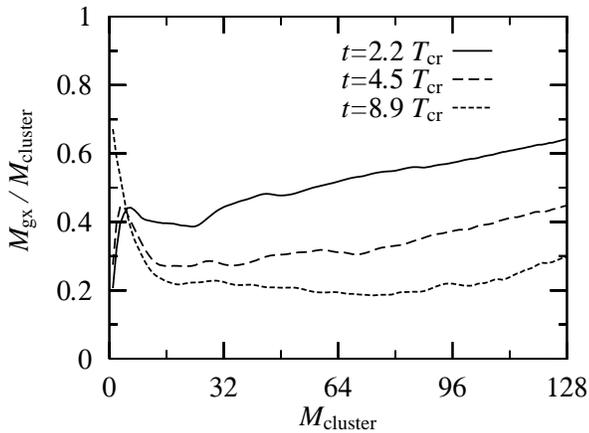}
\caption{%
Fractional mass of the galaxies $M_{\mathrm{gx}}/M_{\mathrm{cluster}}$ plotted against the mass coordinate of the cluster for the run PP.
Solid, long-dashed, and short-dashed curves correspond to $t=2.2\, T_{\mathrm{cr}}$, $4.5\, T_{\mathrm{cr}}$, and $8.9\, T_{\mathrm{cr}}$, respectively.}
\end{figure}

We can get some insights on how the central cusp forms, from the studies of the evolution of star clusters.
When the mass of the stars are all equal, the central density of a star cluster increases through the thermal evolution of the cluster, and the central cusp would develop through the gravothermal catastrophe (Antonov 1962; Lynden-Bell and Wood 1968).
In this case, the slope is around $-2.2$, deeper than the isothermal.
This is because the velocity dispersion goes up in the central region as the result of the gravothermal catastrophe.

When the mass of stars are different, the dynamical friction causes heavy stars to sink toward the center and accelerates the collapse.
In these unequal-mass star clusters, a cusp with a slope shallower than that of the singular isothermal distribution is formed (Inagaki and Wiyanto 1984).
Here, the shallow cusp implies that the velocity dispersion goes down toward the center.
However, the temperature still goes up towards the center, since the average mass goes up towards the center as the result of the mass segregation caused by the dynamical friction.
Apparently, our clusters of galaxies evolve in the same way as the star clusters with stars with different masses.

Takahashi et al.\ (2000) followed the evolution of clusters by solving Fokker--Planck (hereafter FP) equations numerically.
They found the same slope as what we obtained by $N$-body simulation is realized in their FP simulation.
Their result clearly demonstrates that the cusp is formed through the thermal evolution, since only the thermal evolution can occur in FP calculations.

% subsection 4.3
\subsection{Evolution of Galaxies}

In figure~9, the velocity dispersions of galaxies are plotted against their masses for runs PP, PK7 and PH at $t=4.5\, T_{\mathrm{cr}}$.
The velocity dispersions for galaxies are shifted downwards by a factor of two for the run PK7 and by a factor of four for the run PH to make it easy to distinguish them from those of the run PP.
Figure~9 shows that the $m$--$\sigma$ relation does not depend on the initial cluster model and that the galaxies in a cluster evolve along a line $\sigma\propto m^{1/3\sim 1/4}$.

Figure~10 is the same as figure~9, but for runs PP, K9P, and HP.
It is remarkable that for runs K9P and HP, the velocity dispersions were almost constant for the galaxies with $m\gsim 0.4$ (K9P) or $m\gsim 0.5$ (HP), and for those smaller than those conditions, they are distributed along the lines $\sigma\propto m^{1/3\sim 1/4}$.

\begin{figure}[t]
\includegraphics*[width=.48\textwidth]{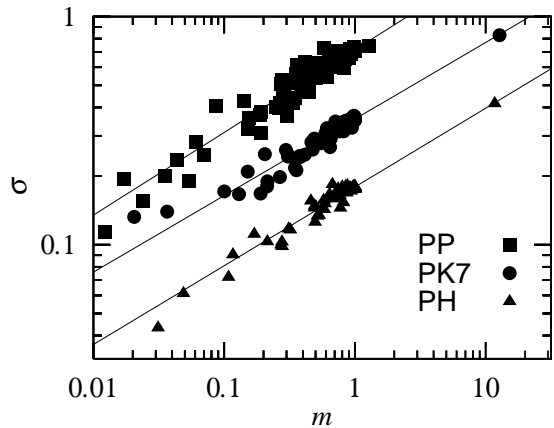}
\caption{%
Velocity dispersions of galaxies plotted against their masses at $t=4.5\, T_{\mathrm{cr}}$ for runs PP (squares), PK7 (circles), and PH (triangles).
Velocity dispersions are shifted by a factor of two for the run PK7, and by a factor of four for the run PH.
The lines are the best-fit lines by least-square fitting.}
\end{figure}

\begin{figure}[t]
\includegraphics*[width=.48\textwidth]{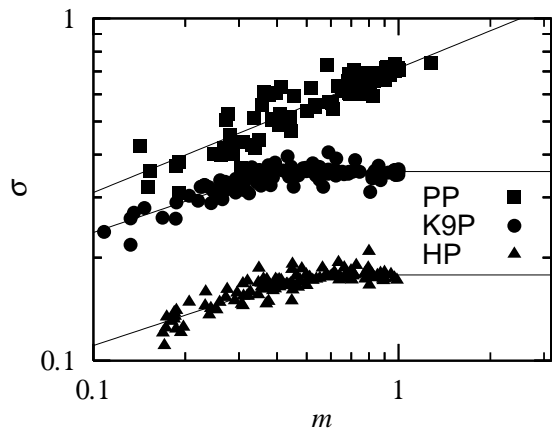}
\caption{%
Same as figure~9, but for runs PP (squares), K9P (circles), and HP(triangles).
Velocity dispersions are shifted by a factor of two for the run K9P, and by a factor of four for the run HP.
For runs K9P and HP, only the galaxies with $m<0.4$ (K9P) or $m<0.5$ (HP) are used to fit.}
\end{figure}

This sharp contrast between centrally concentrated models and less concentrated models are understood as follows.
How the mass and the velocity dispersion of a galaxy change through encounters with another galaxy depends on the structure of the galaxy.
The difference in evolution of the mass--velocity dispersion relation of three galaxy models in figure~10, therefore, should be the result of the difference in initial structures of galaxy models.

Figure~11 shows examples of density profiles of a galaxy for runs PP and K9P.
We chose a galaxy with $m_{\mathrm{gx}}\sim 0.5$ as a sample galaxy for each run at $t=4.5\, T_{\mathrm{cr}}$.
From figure~11(a), we can see that for the case of the Plummer model, the central density decreases as the mass decreases.
On the other hand, for the case of the King~9 model (figure~11(b)), the decrease of density is limited to the outer region.
It means that, for a King model galaxy, the kinetic structure in its inner region is self-gravitating itself so as to change slowly in response to encounters with other galaxies.

Initially both Plummer and King~9 model galaxies have the same virial radii.
However, due to the difference in the central concentration between them, the half-mass radii are different.
The initial half-mass radius of the Plummer model is 0.73 and that of the King~9 model is about 1 (shown by downward arrows in figure~11).

\begin{figure}[t]
(a)
\includegraphics*[width=.48\textwidth]{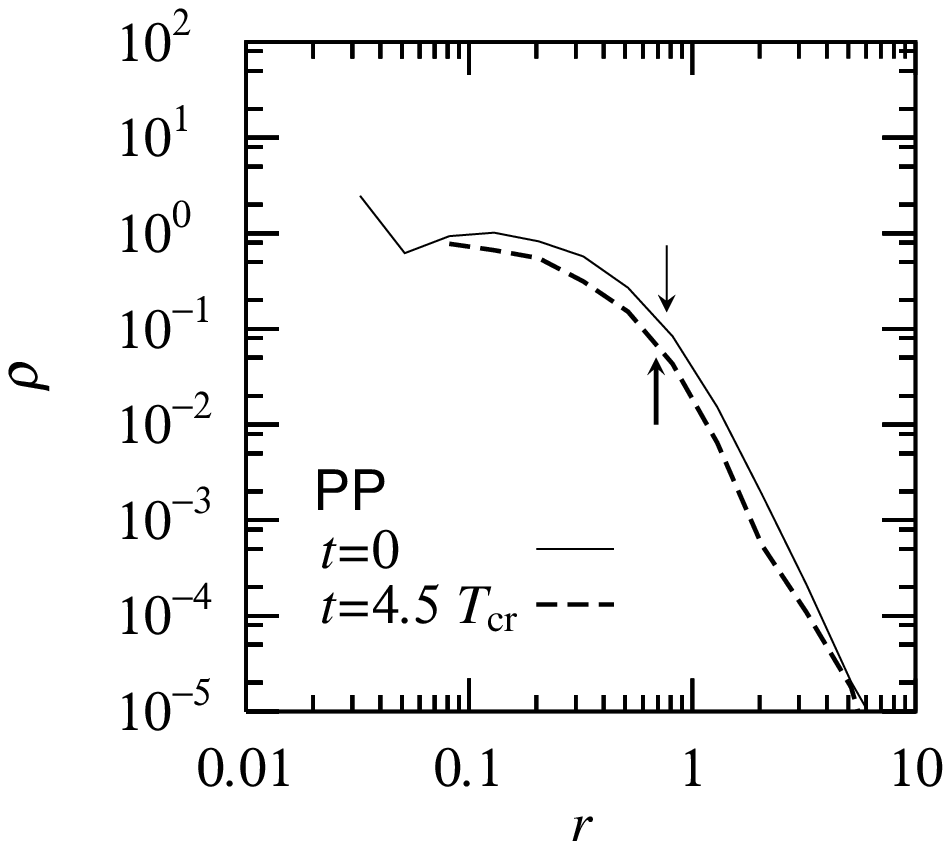}\\[-.4ex]
(b)
\includegraphics*[width=.48\textwidth]{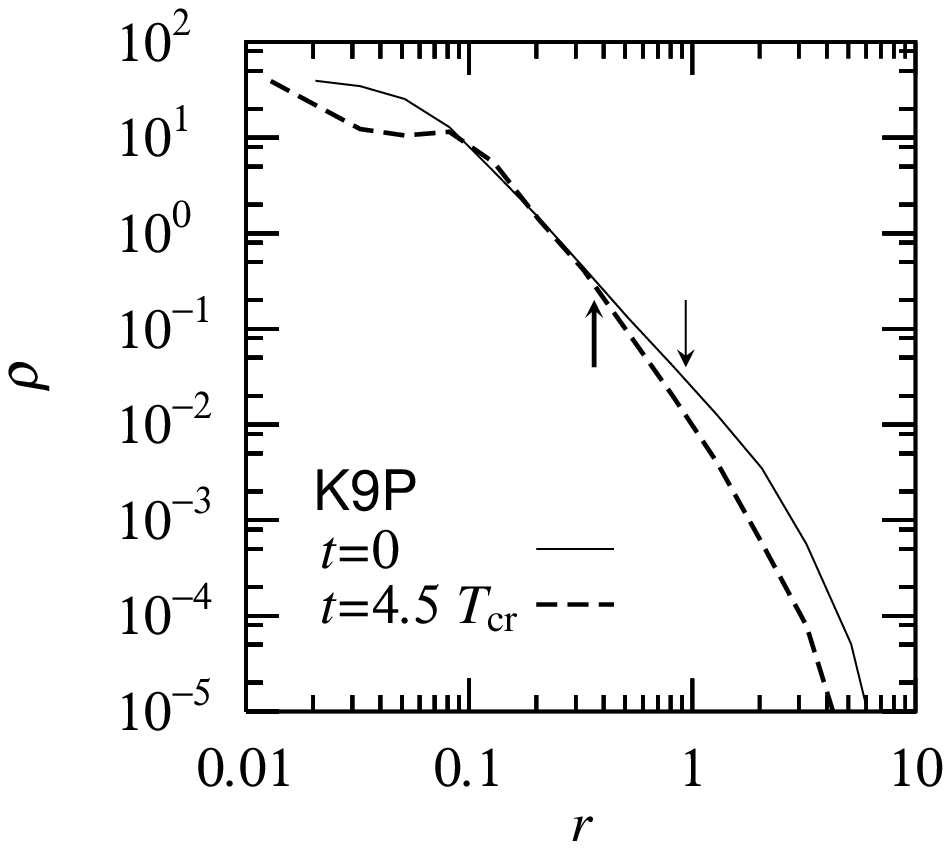}\\[-.4ex]
\caption{%
Density profiles of a galaxy in (a) the run PP and (b) the run K9P.
Solid curves are at $t=0$, and dashed curves are at $t=4.5\, T_{\mathrm{cr}}$.
In each panel, the downward arrow indicates initial half-mass radius of the galaxy, while the upward arrow indicates half-mass radius of the galaxy at $t=4.5\, T_{\mathrm{cr}}$.}
\end{figure}

From these differences, the specific binding energy of a halo star is smaller for the King~9 model than for the Plummer model, if we compare them at the same mass coordinate.
In other words, we can remove large mass from the King~9 model with very little energy input, as far as we are taking out the outer region (say, outside the half-mass radius).
Even the particles only slightly heated up are so easily escaped from the galaxy that the remaining particles do not change their energy much.
Thus, the change in the velocity dispersion is initially small.

On the other hand, to remove stars from the Plummer model, we need to supply much larger energy input even in the outer region.
Since a significant fraction of this energy is spent to heat up the stars which will remain bound (Funato and Makino 1999), the galaxy as a whole expands in the case of the Plummer model, resulting in the decrease in both the velocity dispersion and the central density.

In fact, for a Plummer model galaxy, figure~11(a) also shows that the steep cut-off of density profile in outer region disappeared and a moderately extended halo develops so as to be expressed as $\rho \propto r^{-4}$.
It means that, when the galaxy model was heated up, some particles escape to the intracluster space, while many other heated particles still remain in the galaxy.
Therefore, its specific kinetic energy becomes small and its velocity dispersion decreased.

After around 50\% of its mass is removed, even a King~9 model galaxy no longer has an extended halo.
Thus, further removal of its mass is associated with the net heating, resulting in the decrease in the velocity dispersion.

According to the above argument, the Hernquist model should behave similarly to the King~9 model, since it also has an extended halo.
In fact, as we can see from figure~10, their evolution tracks are very similar.

To summarize, no matter what is the initial profile of the galaxies, their dynamical evolution driven by the encounters with other galaxies follows the direction $\sigma\propto m^{1/3\sim 1/4}$, at least after a significant fraction of the initial mass has been removed from them.

Note that this evolutionary track is consistent with the observed Faber--Jackson relation (Faber and Jackson 1976, hereafter FJ relation) between the luminosity $L$ and velocity dispersion $\sigma$ of cluster ellipticals.
The FJ relation is expressed as $L\propto\sigma^{4}$.
Of course, we need to assume that $M/L$ is constant to relate our numerical result on the relation between the mass and the velocity dispersion and FJ relation between the luminosity and the velocity dispersion.
This might sound unreasonable, since the outskirts of the galaxy must be dominated by the dark matter.
However, as we have seen in figure~10, while the mass in the outskirts is removed, the velocity dispersion remain almost constant.
Therefore, we can conclude that the dynamical evolution of galaxies through encounters is consistent with the observed FJ relation.
We will make more detailed assessment on this subject in section~5.2.

\begin{figure}[t]
\includegraphics*[width=.48\textwidth]{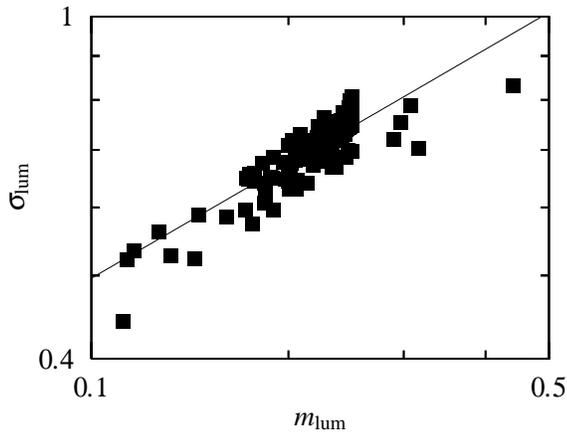}
\caption{%
Velocity dispersions of luminous matters in galaxies plotted against their masses at $t=4.5\, T_{\mathrm{cr}}$ for the run K9P.}
\end{figure}

%
% Summary and Discussion
%

% section 5
\section{Summary and Discussion}

We investigated the evolution of clusters of galaxies using self-consistent $N$-body simulations.
We used different models for a galaxy and a cluster to see how the results depend on the initial model.

In all runs with different galaxy models and different cluster models, we found that about half of the total mass of a cluster escapes from individual galaxies to the intracluster space by about $4\, T_{\mathrm{cr}}$.
This escaped matter forms a common halo.
However, while the cluster is young, the growth rate of the common halo strongly depends on the initial cluster model.
The dependence on the galaxy model is somewhat weaker.

The density profile of the evolved cluster has an approximately $r^{-1.2}$ cusp at the core region for a Plummer model cluster.
The size of this cusp region is roughly the same as the core radius of the initial cluster model.

We also found that the evolution track of individual galaxies in the mass--velocity dispersion plane is consistent with the FJ relation for all galaxy models, after a fair fraction of the total mass is removed.
For models with extended halos such as the King~9 model and the Hernquist model, the velocity dispersion remains almost constant until about half of the total mass escaped to the cluster.

In the following, we first discuss the relation between the initial growth rate of common halos and initial density profiles of cluster models.
Then we discuss the evolution of individual galaxies in connection with FJ relation.

% subsection 5.1
\subsection{Growth Rate of Common Halos}

As shown in section~4.2, the initial growth rate of the common halo depends rather strongly on the initial density profile of the cluster.
Here, we give (semi-)quantitative analytic interpretation on this difference.

In paper I, we showed that the mass-loss rate from individual galaxy was expressed as
%
% equation 10
\begin{equation}
 \frac{dm}{dt} \propto m^{5/2} n r_{\mathrm{h}}^{1/2} V_{\mathrm{c}}^{-2}.
\end{equation}
The growth rate of the common halo is given by integrating the above formula over all galaxies.
Here, let us concentrate on the early phase of the evolution where we can regard $r_{\mathrm{h}}$ as constant.

When we change the cluster model, we can ignore the difference in $V_{\mathrm{c}}$, since all cluster models have the same velocity dispersion.
Therefore, the growth rate of the common halo, $dM/dt$, is simply proportional to the number density of galaxies $n_{\mathrm{gx}}$ integrated over all galaxies.
In other words, we have
%
% equation 11
\begin{equation}
 \frac{dM}{dt} \propto \int d^{3}x\, n_{\mathrm{gx}}^{2}.
\end{equation}
This form is the same as the emission measure for the total X-ray luminosity of the cluster, if we replace $n_{\mathrm{gx}}$ by the electron number density $n_{\mathrm{e}}$.
In both case, we essentially count the number of collisions per unit time.

The initial value of $\displaystyle\int d^{3}x\, n_{\mathrm{gx}}^{2}$ is 0.44 for the cluster of the Plummer model, and 0.82 for the cluster of the King~7 model, which means that the initial growth rate for a King~7 cluster is about twice as large as that for a Plummer cluster.
Table~3 shows the mean growth rates of common halos ($dM/dt$) for runs PP and PK7 in $t\sim 1\, T_{\mathrm{cr}}$, as well as the initial theoretical values.
We can see that the agreement is excellent.

\begin{table}[t]\small
 \caption{Growth rate of common halos in $t\sim 1\,T_{\rm cr}$ (also see figure~3).}
 \begin{center}
  \begin{tabular}{lcc}
   \hline\hline
   \multicolumn{1}{c}{ID} & $dM/dt$ & $\displaystyle\int d^{3}x\, n_{\rm gx}^{2}$ \\
   \hline
   PP  & 0.00678 & 0.44 \\
   PK7 & 0.0147  & 0.82 \\
   \hline
  \end{tabular}
 \end{center}
\end{table}

In a few crossing times of the cluster, many galaxies have disrupted, and many others lose large fraction of their initial masses.
As a result, the growth rate of the common halo in the cluster drops.
This drop is faster for cluster models with high central concentration because of the following two reasons.
First, the galaxies in the central high density region have lost most of the mass.
Therefore there is not much mass left to escape.
Second, these cluster models also have extended halos, which means a fair fraction of galaxies are in very low density region.
These galaxies cannot lose much mass.
These combined effect explains why the growth rate of the common halo in the late phase is slower for models with high central density.

% subsection 5.2
\subsection{$m$--$\sigma$ Relation of Individual Galaxies}

As we have seen in section~4.3, the relation $\sigma\propto m^{1/3\sim 1/4}$ develops for a cluster with Plummer model galaxies as galaxies evolve.
In the case of clusters with King~9 or Hernquist model galaxies, the same relation also develops after galaxies lost about more than half of their masses.

In the following, we discuss how the FJ relation is realized in observed clusters of galaxies, under the standard CDM model and the hierarchical structure formation scenario.

When galaxies are initially formed, they should have the dark matter more extended than the luminous matter, since the dark matter is dissipationless while the baryonic matter, from which the stars will eventually be formed, is dissipational.
Recent high-resolution simulations of formation of dark-matter halos (Fukushige and Makino 1997, Moore et al.\ 1998, Fukushige and Makino 2000) suggest that the dark-matter halos have the central cusp slightly shallower than isothermal ($\rho \propto r^{-1.5}$) which smoothly connect to the outer halo with $\rho \propto r^{-3}$.
In other words, they have a rather extended (nearly) isothermal region.

In the cluster environment, the outer halo is quickly stripped through encounters with other galaxies to form a steeper halo with $\rho \propto r^{-4}$, as we have seen in the case of the King~9 model.
During this period, however, the luminosity and the velocity dispersion of the luminous matter would suffer very little change, since the luminous matter is more centrally concentrated.

When the luminous matter starts to be stripped through encounters, the morphology of the galaxy itself starts to change, since the stellar disk would be heated up by encounters.
As a result, they would become E/S0 galaxies from spirals.
Once the galaxy reached this regime, we can regard its $M/L$ to be roughly constant, since the luminous matter bears a large fraction of the total mass anyway.
Therefore, further stripping would result in the relation $\sigma\propto m^{1/4}$, which can be regarded as the FJ relation.

To see what we would see for the true $L-\sigma$ relation, in figure~12 we plot the relation between the ``luminosity'' and the velocity dispersion for the run PK9.
Here, we simply assumed that particles which were within 25\% mass radius at $t=0$ are all luminous matters, and remaining 75\% is dark.
We see that the slope is rather steep when the mass (``luminosity'') is still large, but it becomes shallower to approach $-1/4$ as the mass becomes smaller.
Best fit for the result shown in figure~12, $\sigma_{\mathrm{lum}} \propto m_{\mathrm{lum}}^{0.43}$, is significantly steeper than the observed FJ relation.

Qualitatively, this implies that our scenario described above works fine.
By limiting the luminous matter to the central region, we eliminated the region in figure~11(b) where the velocity dispersion is almost constant.
Unfortunately, our simple model is a bit too successful, resulting in the relation too steep when the mass is still large.

The reason why we obtained steeper relation in figure~12 is probably that our model for the luminous matter has a cutoff which is too sharp.
Since the cutoff is sharp, the luminosity cannot change until the stripping reacheppps to the very central region, while the velocity dispersion starts to decrease when dark matters just outside the luminous matter are removed.
We need to employ more realistic model for the distribution of the luminous matter.
However, at least qualitatively, we can conclude that the mass loss through encounters is consistent with the observed FJ relation.

We are not arguing that the mass loss through encounters is the only mechanism to produce the FJ relation.
The merging of two galaxies also drives the evolution of galaxies along the line of $m\propto \sigma^4$ on $m-\sigma $ plane (Farouki et al.\ 1983).
Hierarchical merging predicted in CDM scenario, therefore, would also result in the $m-\sigma$ relation which is consistent with the FJ relation.

The important point is that the dynamical evolution and interaction between galaxies would not break the FJ relation even if the dynamical environment is so violent.
Our result shows that the FJ relation is very robust relation against dynamical evolution of galaxies.

\vspace{1pc}

This work was supported by Research for the Future Program of the Japan Society for the Promotion of Science, JSPS-RFTP 97P01102.

%
% References
%

\section*{References}

\re
Aarseth S.\ J., Fall S.\ M.\ 1980, ApJ 236, 43
\re
Antonov, V.\ A.\ 1962, Solution of the problem of stability of stellar system Emden's density law and the spherical distribution of velocities (Vestnik Leningradskogo Universiteta, Leningrad: University)
\re
Athanassoula E., Bosma A., Lambert J.-C., Makino J.\ 1998, MNRAS 293, 369
\re
Athanassoula E., Makino J., Bosma A.\ 1997, MNRAS 286, 825
\re
Barnes J.\ E., Hut P.\ 1986, Nature 324, 446
\re
Bertschinger E.\ 1998, ARA\&A 36, 599
\re
Bode P.\ W., Berrington R.\ C., Cohn H.\ N., Lugger P.\ M.\ 1994, ApJ 433, 479
\re
Byun Y.-I., Grillmair C.\ J., Faber S.\ M., Ajhar E.\ A., Dressler A., Kormendy J., Lauer T.\ R., Richstone D., Tremaine S.\ 1996, AJ 111, 1889
\re
Casertano S., Hut P.\ 1985, ApJ 298, 80
\re
Couch W.\ J., Barger A.\ J., Smail I., Ellis R.\ S., Sharples R.\ M.\ 1998, ApJ 497, 188
\re
Dressler A.\ 1980, ApJ 236, 351
\re
Faber S.\ M., Jackson R.\ E.\ 1976, ApJ 204, 668
\re
Fall S.\ M.\ 1978, MNRAS 185, 165
\re
Farouki, R.\ T., Shapiro, S.\ L., Duncan, M.\ J.\ 1983, ApJ 265, 597
\re
Fukushige T., Makino J.\ 1997, ApJ 477, L9
\re
Fukushige T., Makino J.\ submitted (\texttt{astro-ph/0008104})
\re
Funato Y., Makino J.\ 1999, ApJ 511, 625
\re
Funato Y., Makino J., Ebisuzaki T.\ 1993, PASJ 45, 289
\re
Garijo A., Athanassoula E., Garc{\'{\i}}a-G{\'{o}}mez C.\ 1997, A\&A 327, 930
\re
Ghigna S., Moore B., Governato F., Lake G., Quinn T., Stadel J.\ 1998, MNRAS 300, 146
\re
Heggie D.\ C., Mathieu R.\ D.\ 1986, in The Use of Supercomputers in Stellar Dynamics, ed P.\ Hut, S.\ McMillan (Springer, Berlin) p233
\re
Hernquist L.\ 1990, ApJ 356, 359
\re
Inagaki, S., Wiyanto, P.\ 1984, PASJ 36, 391
\re
Kawai A., Fukushige T., Makino J., Taiji M.\ 2000, PASJ 52, 659
\re
King I.\ R.\ 1966, AJ 71, 64
\re
Lauer T.\ R., Ajhar E.\ A., Byun Y.-I., Dressler A., Faber S.\ M., Grillmair C., Kormendy J., Richstone D., Tremaine S.\ 1995, AJ 110, 2622
\re
Lynden-Bell, D., Wood, R.\ 1968, MNRAS 138, 495
\re
Makino J.\ 1991, PASJ 43, 621
\re
Makino J., Taiji M., Ebisuzaki T., Sugimoto D.\ 1997, ApJ 480, 432
\re
Miyoshi K., Kihara T.\ 1975, PASJ 27, 333
\re
Moore B., Governato F., Quinn T., Stadel J., Lake G.\ 1998, ApJ 499, L5
\re
Navarro J.\ F., Frenk C.\ S., White S.\ D.\ M.\ 1997, ApJ 490, 493
\re
Okamoto T., Habe A.\ 1999, ApJ 516, 591
\re
Sensui T., Funato Y., Makino J.\ 1999, PASJ 51, 943 (paper I)
\re
Takahashi K., Sensui T., Funato Y., Makino J.\ 2000, preprint
\re
van Dokkum P.\ G., Franx M., Fabricant D., Kelson D.\ D., Illingworth G.\ D.\ 1999, ApJ 520, L95

\label{last}

\end{document}